\documentclass[
    ,final            
  ]
  {aipproc}

\layoutstyle{6x9}

\newcommand{\be}{\begin{equation}}
\newcommand{\ee}{\end{equation}}
\newcommand{\ba}{\begin{eqnarray}}
\newcommand{\ea}{\end{eqnarray}}

\begin{document}

\title{Turbulence Spectra from Doppler-shifted Spectral Lines}


\author{A. Lazarian}{
  address={Department of Astronomy, University of Wisconsin-Madison, lazarian@astro.wisc.edu}
}

\begin{abstract}
 Turbulence is a key element of the dynamics of astrophysical
fluids, including those of interstellar medium, clusters of galaxies and 
circumstellar regions. Turbulent motions induce Doppler
shifts of observable emission and absorption lines. In the
review we discuss new techniques that relate the
spectra of underlying velocity turbulence and spectra of Doppler-shifted lines.
  In particular, the Velocity-Channel
Analysis (VCA) makes use of the channel maps, while the Velocity Coordinate Spectrum
(VCS) utilizes the fluctuations measured along the velocity axis of the  Position-Position
Velocity (PPV) data cubes.
Both techniques have solid foundations based on analytical calculations as well
as on numerical testings. 
Among the two the VCS, which has been developed quite recently,
has two advantages. First of all, it is applicable to turbulent volumes that
are not spatially resolved. Second, it can be used with absorption lines that
do not provide good spatial sampling of different lags over the image of turbulent object.
In fact, numerical testing shows that
 measurements of Doppler shifted absorption lines over a few directions is sufficient
for a reliable recovering of the underlying spectrum of the turbulence.
Our comparison of the VCA and the VCS with a more traditional technique of Velocity Centroids,
shows that the former two techniques recover reliably the spectra of supersonic turbulence,
while the Velocity Centroids may have advantages for studying subsonic turbulence.
In parallel with theoretical and numerical work on the VCA and the VCS,
the techniques have been applied to spectroscopic observations. 
We discuss results on astrophysical turbulence obtained with the VCA and the VCS.
\end{abstract}

\maketitle


\section{What can Turbulent Spectra Tell us?}

 As a rule astrophysical fluids are turbulent and the turbulence 
is magnetized. This ubiquitous turbulence determines the transport
properties of interstellar medium (see Elmegreen \& Falgarone 1996, Stutzki 2001,
Balesteros-Peredes et al. 2006) and intracluster medium
(see Sunyaev, Norman \& Bryan 2003, Ensslin \& Vogt 2006, Lazarian 2006), 
many properties of Solar and stellar
winds (see Hartman \& McGregor 1980) etc. One may say that to understand heat 
conduction, propagation
of cosmic rays and electromagnetic radiation in different astrophysical environments it
is absolutely essential to understand the properties of underlying magnetized turbulence.
The fascinating processes of star formation (see McKee \& Tan 2002, 
Elmegreen 2002,  Mac Low \& Klessen 2004) and interstellar chemistry (
see Falgarone et al. 2006 and references therein)
are also intimately related
to properties of magnetized compressible turbulence (see reviews by
Elmegreen \& Scalo 2004). 

From the point
of view of fluid mechanics astrophysical turbulence 
is characterized by huge Reynolds numbers, $Re$, which is  
the inverse ratio of the
eddy turnover time of a parcel of gas to the time required for viscous
forces to slow it appreciably. For $Re\gg 100$ we expect gas to be
turbulent and this is exactly what we observe in HI (for HI $Re\sim 10^8$).
In fact, very high astrophysical $Re$ and its magnetic counterpart
magnetic Reynolds number $Rm$
 (that can be as high as $Rm\sim 10^{16}$) present a big problem for numerical simulations
that cannot possibly get even close to the astrophysically-motivated numbers. The currently available 3D
simulations can have $Re$ and $Rm$ up to $\sim 10^4$.  Both scale as
the size of the box to the first power, while the computational effort
increases as the fourth power (3 coordinates + time), so the brute force 
approach cannot begin to resolve the controversies related to ISM turbulence.
This caused serious concerns that while present codes can produce simulations that
resemble observations, whether numerical simulations reproduce reality well 
(see McKee 1999, Shu et al. 2004). We believe that these concerns may be addressed via
observational studies of astrophysical turbulence.

Statistical description is a nearly indispensable strategy when
dealing with turbulence. The big advantage of statistical techniques
is that they extract underlying regularities of the flow and reject
incidental details. Kolmogorov description of unmagnetized incompressible
turbulence is a statistical
one. For instance it predicts that the difference in velocities at
different points in turbulent fluid increases on average
with the separation between points as a cubic root of the separation,
i.e. $|\delta v| \sim l^{1/3}$. In terms of direction-averaged
energy spectrum this gives the famous Kolmogorov
scaling $E(k)\sim 4\pi k^2 P({\bf k})\sim k^{5/3}$, where $P({\bf k})$ 
is a {\it 3D} energy spectrum defined as the Fourier transform of the
correlation function of velocity fluctuations $\xi ({\bf r})=\langle  
\delta v({\bf x})\delta v({\bf x}+{\bf r})\rangle$. Note that in
this paper we use $\langle  ...\rangle$ to denote averaging procedure.

The example above shows the advantages of the statistical approach
to turbulence. For instance, the energy spectrum 
$E(k)dk$ characterizes how much
energy resides at the interval of scales $k, k+dk$. At large scales $l$
which correspond to small wavenumbers $k$ ( i.e. $l\sim 1/k$) one expects
to observe features reflecting energy injection. At small scales
one should see the scales corresponding to
sinks of energy. In general, the shape of the spectrum is
determined by a complex process of non-linear energy transfer and
dissipation. 

In view of the above it is not surprising that attempts
to obtain spectra of interstellar turbulence have been numerous since 1950s
(see Munch 1958). However, various directions
of research achieved various degree of success (see 
Armstrong, Rickett
\& Spangler 1995). 
For instance, studies of turbulence statistics of ionized media 
were more successful
(see Spangler \& Gwinn 1990) and provided the information of
the statistics of plasma density at scales $10^{8}$-$10^{15}$~cm. 
However, these sort of measurements provide only the density statistics, 
which is an indirect measure of turbulence. 

Velocity statistics is much more coveted turbulence measure. 
Although, it is clear that Doppler broadened lines
are affected by turbulence, recovering of velocity statistics was
extremely challenging without an adequate theoretical insight.
Indeed, both velocity
and density contribute to fluctuations of the intensity in the 
Position-Position-Velocity (PPV) space. In what follows we discuss how
the observable Doppler-shifted lines can be used to recover a spectrum of
turbulent velocity using two new techniques, that, unlike other mostly empirical
techniques, have solid theoretical foundations.
 How to obtain using spectroscopic observations other characteristics of turbulence,
e.g. higher order statistics, anisotropies
has been reviewed earlier (see Lazarian 2004). 

\section{VCA and VCS: Two Ways to Analyze Spectral Data}

Spatial spectra obtained by taking Fourier transform  of channel maps had been widely used to study HI 
before we conducted our
 theoretical study of what those spectra mean in Lazarian \& Pogosyan (2000, henceforth LP00). 
The channel maps correspond to the velocity slices of PPV cubes as shown in 
Figure~1 and one may naturally ask a question whether anything depends on the
thickness of the channel. It is intuitively clear that if medium is optically thin and the 
velocity is integrated over the whole spectral line,
the fluctuations can depend only on density inhomogeneities. It is also suggestive that 
the contribution
of the velocity fluctuations may depend on whether the images of the eddies under study 
fit within a
velocity slice or their velocity extend is larger than the slice thickness (see Figure~1).
 According to LP00 this results in
the assymptotics that correspond to ``thin'' and ``thick'' velocity channels. Note, 
that these questions
were not posed by the earlier research. This resulted in channel map spectra with different spectral
indexes the relation of which to the underlying velocity fluctuations
 was unclear (see Green 1990 and references 
therein).


\begin{figure}
\hbox{
  \includegraphics[height=.3\textheight]{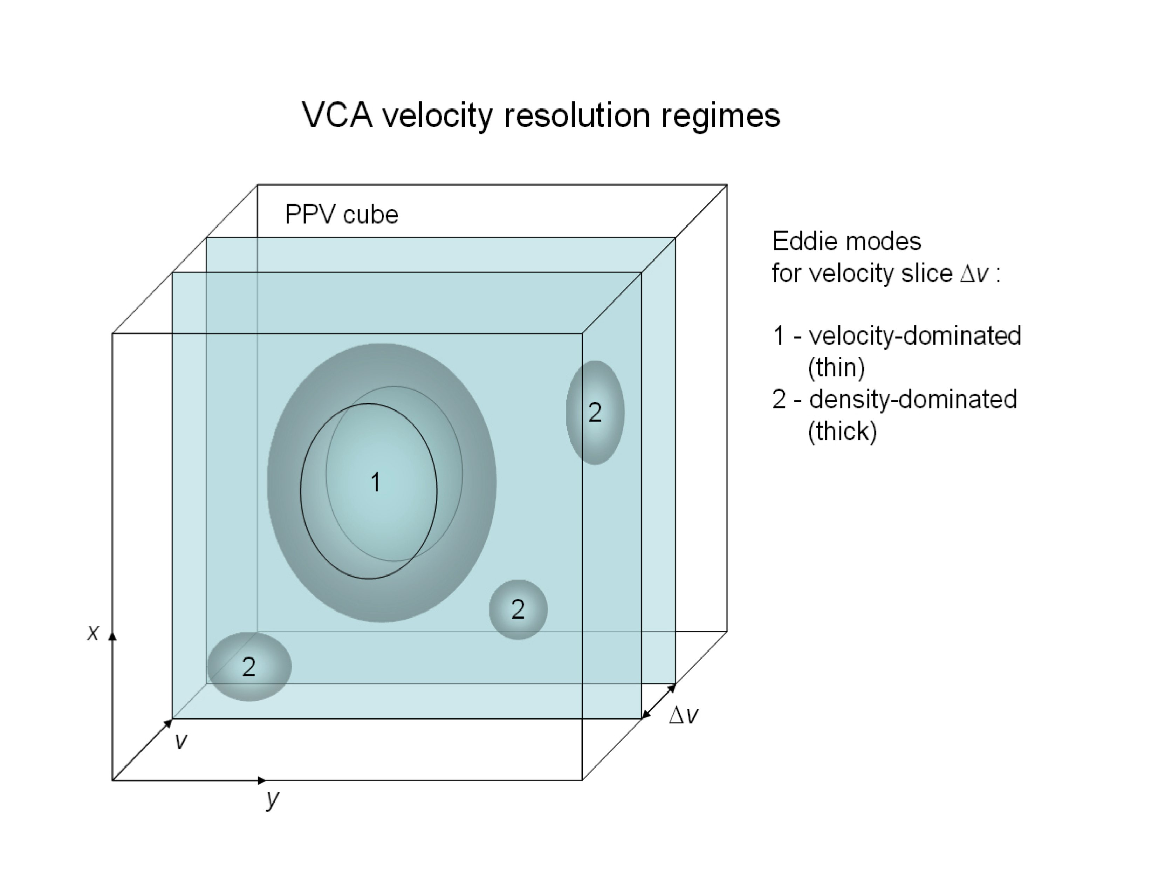}
\hfill
 \includegraphics[height=.3\textheight]{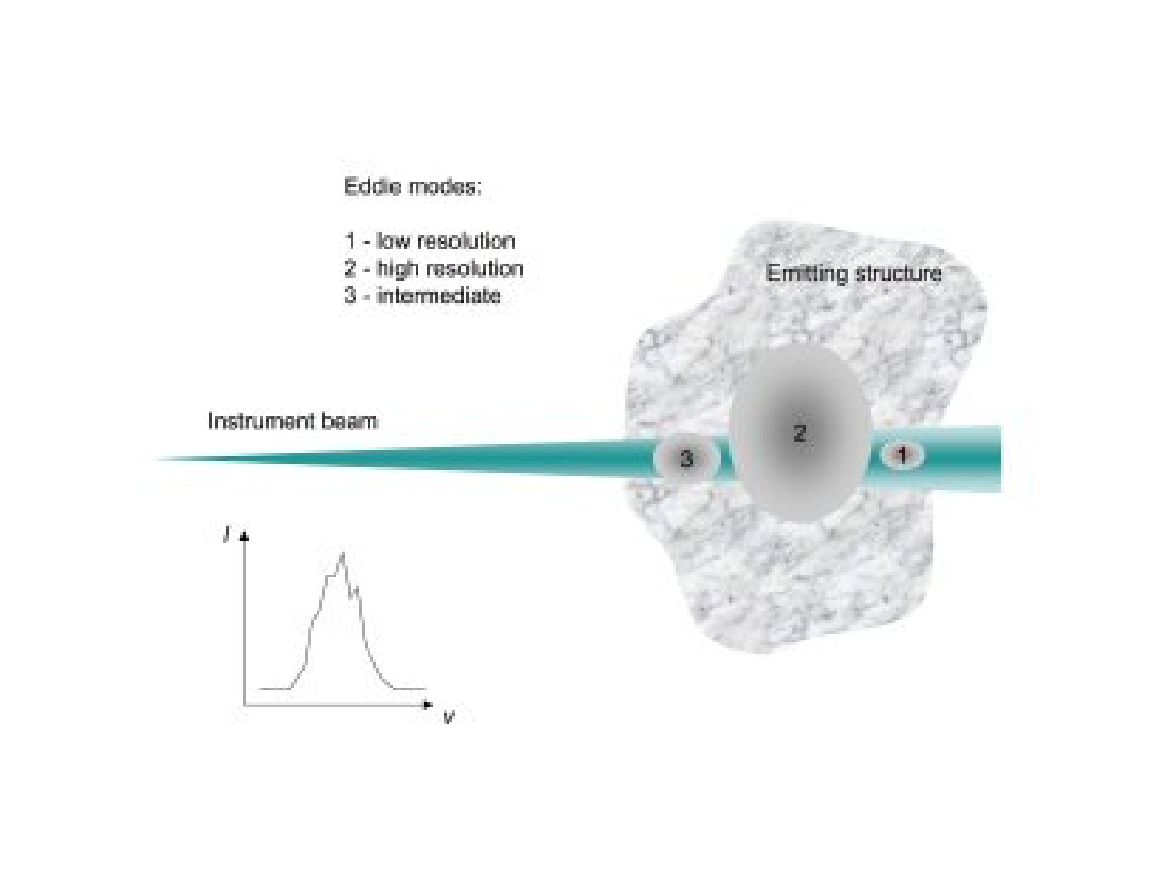}
}
  \caption{{\it Left Panel}: PPV data cube. Illustration of the concepts of the thick and thin velocity
slices. the slices are thin for the PPV images of the large eddies, but thick for the images of small
eddies. {\it Right Panel}: Illustration of the VCS technique. For a given instrument resolution large
eddies are in the high resolution limit, while small eddies are in the low resolution limit.}
\end{figure}

Some spectral data, for instance, optically thick 
CO data were traditionally analyzed in differently (see Falgarone \& Puget 1995), namely, spectra
of total intensities was studied. The origin of this spectrum and its relation
to the underlying velocity and density fluctuations was established in 
Lazarian \& Pogosyan (2004, henceforth LP04).
This work also clarified the effects of absorption that were reported for HI data. 
In terms of the techniques of turbulence study LP04 deals with the
VCA, but in the case when absorption is present.

A radically different way of analyzing spectroscopic
 data is presented in Lazarian \& Pogosyan (2006, henceforth LP06) and Chepurnov \& Lazarian (2006a).
There the spectra along the V-axis of the PPV cube are studied (see Figure~2).
 The technique was termed Velocity
Coordinate Spectrum (VCS) in Lazarian (2004). The formalism can be traced to LP00 work, where the
expressions that relate the spectrum of fluctuations along the velocity coordinate and the underlying
velocity spectrum were obtained. However, it took some time to understand the advantages that the 
VCS provides for the practical handling of the emission and absorption data.
 The first analysis of the data using the VCS is performed
by Chepurnov \& Lazarian (in preparation). Numerical testing of the technique is provided
in Chepurnov \& Lazarian (2006b).

\section{Basics of the Formalism}

Below we provide a brief introduction to the mathematical foundations of the VCA and the VCS
(see more in
LP00, LP04 and LP06).
Our goal is to relate the statistics
that can be obtained through spectral line observations, for instance,
the structure function of the intensity of emission $I_{\bf X}(v)$ 
\begin{equation}
{\cal D}({\bf X}, v_1,v_2)\equiv \left\langle \left[ I_{\bf X}(v_1) -
I_{\bf X}(v_2)\right]^2\right\rangle~~~,
\label{dv_emiss}
\end{equation}
where the $z$-axis component velocity
$v$ is measured by in the direction defined by the two dimensional
vector\footnote{Henceforth we denote by the capital bold letters
the two dimensional position-position vectors that specify the line of sight.
Small bold letters are reserved to describe the vectors
of three dimensional spatial position.  The $z$-axis 
is chosen to be along the line of sight.}
${\bf X}$, to the underlying properties of the turbulent cascade.

The intensities $I_{\bf X}(v)$ are affected by both the turbulence and
the absorption. To quantify these effects we
consider the standard equation of radiative transfer
$
dI_{\nu}=-g_{\nu} I_{\nu} ds+j_{\nu}ds~~~,
$
where 
$g_{\nu}=\alpha({\bf x}) \rho({\bf x}) \phi_v({\bf x})$,
$j_{\nu}=\tilde{\epsilon} \rho({\bf x}) \phi_v({\bf x})$, ${\bf x}$ is a three
dimensional position vector $({\bf X}, z)$, $\rho({\bf x)}$ is the density and 
$\phi_v({\bf x})$ is the 
velocity distribution of the atoms. The turbulent motions affect
the velocity distribution. Indeed,
for the line-of-sight component, $v$, of the velocity 
at the position ${\bf x}$, it is a sum of the $z$-components of the
regular gas flow (e.g., due to galactic rotation) $v_{gal}({\bf x})$,
the turbulent velocity $u({\bf x})$ and the residual component
due to thermal motions. This residual thermal
velocity $v-v_{gal}({\bf x})-u({\bf x})$ has a Maxwellian distribution,
so
\begin{equation}
\phi_v({\bf x}) {\mathrm d} v =\frac{1}{(2\pi \beta)^{1/2}}
\exp\left[-\frac{(v-v_{gal}({\bf x})-u({\bf x}))^2}
{2 \beta }\right] {\mathrm d} v ~~~,
\label{phi}
\end{equation}
where $\beta=\kappa_B T /m$, $m$ being the mass of atoms. For $T\rightarrow 0$ the
function $\phi_v$ tends to a delta-function that depends on regular gas flow and
the turbulent velocity $u$. Taking Fourier transforms we deal with velocity
gradients, which are larger for turbulent motions than for large-scale sheer. For instance,
the latter
for the Galactic rotation is given by the Oort's constant, which is $14$~km~s$^{-1}$ kpc$^{-1}$.
In comparison, the shear due to typical Kolmogorov-type
turbulent motions in the Galaxy with the injection
of energy at 10 km~s$^{-1}$ at the scale of $L\sim 30$~pc is $\sim 300$~km~s$^{-1} (L/l)^{2/3}$ kpc$^{-1}$. Thus, in spite of the fact, that
regular large-scale galactic shear velocities may be much larger than the turbulent velocities, 
they can be neglected for our analysis (LP00 and a numerical study in Esquivel et al. 2003). 

For the $z$-component of the turbulent velocity field
(i.e. $u$)  we use the structure function 
\begin{equation}
D_z({\bf r})=
\langle(u({\bf x}+{\bf r})-u({\bf x}))^2 \rangle ~~, 
\label{Dz}
\end{equation}
which for a self-similar power-law turbulent motions
 provide $D_z\sim r^m$, where $m=1/3$ for Kolmogorov turbulence.
These velocity correlations together with the correlations of over-density
\be
\xi(r)=\xi({\bf r}) = \langle \rho ({\bf x}) \rho ({\bf x}+{\bf r}) \rangle~~
~,
\label{xifirst}
\ee
enter the correlation function that can be constructed from the PPV densities $\rho_s$, which are
available through spectroscopic observations.
 If the gas is confined in an isolated cloud of size $S$ and the
galactic shear over this scale is neglected, the zero-temperature correlation
function is (see LP06)
\begin{equation}
\xi_s(R,v)\equiv \langle \rho_s({\bf X_1},v_1)\rho({\bf X_2}, v_2)\rangle \propto
\int_{-S}^S {\mathrm d}z \left(1-\frac{|z|}{S}\right)
\; \frac{\xi( r)}{D_z^{1/2}({\bf r})}
\exp\left[-\frac{v^2}{2 D_z({\bf r})}\right],
\label{ksicloud}
\end{equation}
where the correlation function $\xi_s$ is defined in the PPV space, where $R$ is the spatial separation
between points in the plane-of-sky and $v$ is the separation along the V\
-axis.

\begin{table}[htb]
\begin{tabular}{lccc} \hline\hline\\
& \multicolumn{1}{c}{ 1D: $P_s(k_v)$} & 2D: $P_s(K)$ & 3D: $P_s(K,k_v)$  \\[2mm]
& \multicolumn{1}{c}{ $ k_v D_z^{1/2}(S) \gg 1 $} & $ KS \gg 1 $
&  $ k_v^2 D_z(S) \gg (k S)^m $ 
\\[2mm] \hline \\
$ P_{\rho}: $ & $(r_0/S)^\gamma \left[k_v D_z^{1/2}(S)\right]^{2(\gamma-1)/m} $
& $ \left(r_0/S\right)^\gamma \left[K S\right]^{\gamma+m/2-3} $
&  $ (r_0/S)^\gamma \left[k_v D_z^{1/2}(S)\right]^{-2 (3-\gamma)/m} $ \\[2mm] 
\hline \\
$ P_{v}:$ & $ \left[k_v D_z^{1/2}(S)\right]^{-2/m}$
& $\left[K S\right]^{m/2-3}$
& $\left[k_v D_z^{1/2}(S)\right]^{-6/m} $ \\[3mm]
\hline
\end{tabular}
\caption{The short-wave asymptotical behavior
of power spectra  in PPV space. $\gamma$ is the power-law index of the density correlation
function, $m$ is the index of the velocity correlation function. The spectrum of fluctuations
in the PPV volume can be presented as the sum of $P_{\rho}$, which has both density and
velocity contributions and $P_{v}$, which has only the contribution arising from turbulent
velocity. From LP06.}
\label{tab:1Dspk_asymp}
\end{table}

The correlation function of over-density given by  Eq.~(\ref{xifirst}) has a constant part that depends on the mean density
only and the part that changes with $r$. For instance, for the power-law
density spectrum the correlation functions of over-density take the form (see LP06
for the discussion of cases of $\gamma<0$ and $\gamma>0$):
\begin{equation}
\xi(r)= \langle \rho \rangle^2 
\left(1 + \left[ {r_0 \over r} \right]^\gamma\right), 
\label{Appeq:xi}
\end{equation}
where $r_0$ has the physical meaning of the scale at which fluctuations
are of the order of the mean density (see more in LP06). Substituting Eq.~(\ref{Appeq:xi})
in Eq.~(\ref{ksicloud}) it is easy to see that
 the PPV correlation
function $\xi_s$ can be presented as a sum of two
terms, one of which does depend on the fluctuations of density, the other does not.
Taking Fourier transform of $\xi_s$ one gets the PPV spectrum that evidently is also
a sum of two terms $P_{\rho}$ and $P_{v}$  the assymptotics for which are presented in
Table~1.

Expression (\ref{ksicloud}) and its generalizations
may be used directly to solve the inverse problem to find the properties of the underlying astrophysical
turbulence for an arbitrary spectrum.
 However, so far, most attention was given to astrophysically important case of
power-law turbulence. Note, that using spectra rather than the 
correlation function has advantages. For instance, the correlations along the V-axis of the PPV cube may be
dominated by large-scale gradients, while spectra provide correct result (see
explanation in LP04 and LP06).  The results for 1, 2 and 3 dimensional spectra are presented in
Table~1. In many cases the contributions that depend on density are subdominant. Table~2 presents special
cases when density fluctuations are important.
\begin{table}[t]
\begin{tabular}{llc}
\hline
$ m \ge \mathrm{max}\left[\frac{2}{3},\frac{2}{3}(1-\gamma)\right] $
 & $ v^2 <  D_z(S) (r_0/S)^m $ \\
$ \frac{2}{3}(1-\gamma) < m < \frac{2}{3} $
& $ v^2 < D_z(S) (r_0/S)^{\frac{2/3 \gamma m}{m-2/3(1-\gamma)}} $ \\
$ m \le \mathrm{min}\left[\frac{2}{3},\frac{2}{3}(1-\gamma)\right] $
& $ r_0/S > 1 $ 
\end{tabular}
\caption{Conditions for the impact of
density inhomogeneities to the PPV statistics exceeds
the velocity contribution. $\gamma$ must be larger than 0.}
\label{table:density}
\end{table}

\section{Velocity Channel Analysis}

The interpretation of the channel maps is the domain of the VCA. 
Table~3 shows how the power spectrum of the intensity fluctuations depends on the
thickness of the velocity channel. Below we provide quantitative discussion of the VCA.

\begin{table}
\begin{tabular}{lcc}
\hline
Slice & Shallow 3-D density & Steep 3-D density\\
thickness & $P_{n} \propto k^{-3+\gamma}$, $\gamma>0$ &$P_{n} \propto k^{-3+\gamma}$, $\gamma<0$\\
\hline
2-D intensity spectrum for thin~~slice &
$\propto K^{-3+\gamma+m/2}$    & $\propto
K^{-3+m/2}$   \\
2-D intensity spectrum for thick~~slice & $\propto K^{-3+\gamma}$
& $\propto K^{-3-m/2}$  \\
2-D intensity spectrum for very thick~~slice & $\propto K^{-3+\gamma}$ & $\propto \
K^{-3+\gamma}$  \\
\hline
\end{tabular}
\caption{The VCA assymptotics. {\it Thin} means that the 
channel width $<$ velocity dispersion at the scale under
study;
{\it thick} means that the 
channel width $>$ velocity dispersion at the scale under
study;
{\it very thick} means that a
substantial part of the velocity profile is integrated over.
}
\end{table}

It is easy to see that both steep and shallow underlying density
the power law index
{\it steepens} with the increase of velocity slice
thickness. In the thickest velocity slices the velocity information
is averaged out and we get the
density spectral index $-3+\gamma$. The velocity fluctuations dominate in
thin slices, 
and the index $m$ that characterizes the velocity  fluctuation
can be obtained using thin velocity slices (see Table~1). As we mentioned earlier, the
notion of thin and thick slices depends on a turbulence scale under
study and the same slice can be thick for small scale turbulent fluctuations
and thin for large scale ones (see Figure~1).

One may notice that the spectrum
of intensity in a thin slice gets shallower as the underlying
velocity get steeper. To understand this effect let us consider turbulence
in  an incompressible optically thin medium. The intensity of emission
in a velocity slice is proportional to the number of atoms per
the velocity interval given by the thickness of the slice.
Thin slice means that the velocity dispersion at the scale of
study is larger than the thickness of a slice. The increase
of the velocity dispersion at a particular scales means that
less and less energy is being emitted within the velocity
interval that defines the slice. As the result the PPV image of
the eddy gets fainter. In other words, the larger is the
dispersion at the scale of the study the less intensity
is registered at this scale within the thin slice of spectral
data. This means that steep velocity spectra that correspond
to the flow with more energy at large scales should produce
intensity distribution within thin slice for which the
more brightness will be at small scales. This is exactly
what our formulae predict for thin slices (see Table~3).

 If density variations are
also present they modify the result. 
For small scale asymptotics of thin slices this happens, however, only when the density spectrum
is shallow (i.e. $\gamma>0$), i.e. dominated by fluctuations at small scales (see Eq.~(\ref{Appeq:xi})).

\section{Velocity Coordinate Spectrum}

The VCS is a brand new technique, which, unlike the VCA, was not motivated by the 
interpretation of the
existing observations. In the case of the VCS it were theoretical advances that induced the 
subsequent data analysis.

Unlike the standard spatial spectra, that are functions of angular wavenumber, the VCS is a function of
the wave number $k_v\sim 1/v$, which means that large $k_v$ correspond to small velocity differences,
while small $k_v$ correspond to large velocity differences. 
\begin{figure}
\hbox{
  \includegraphics[height=.3\textheight]{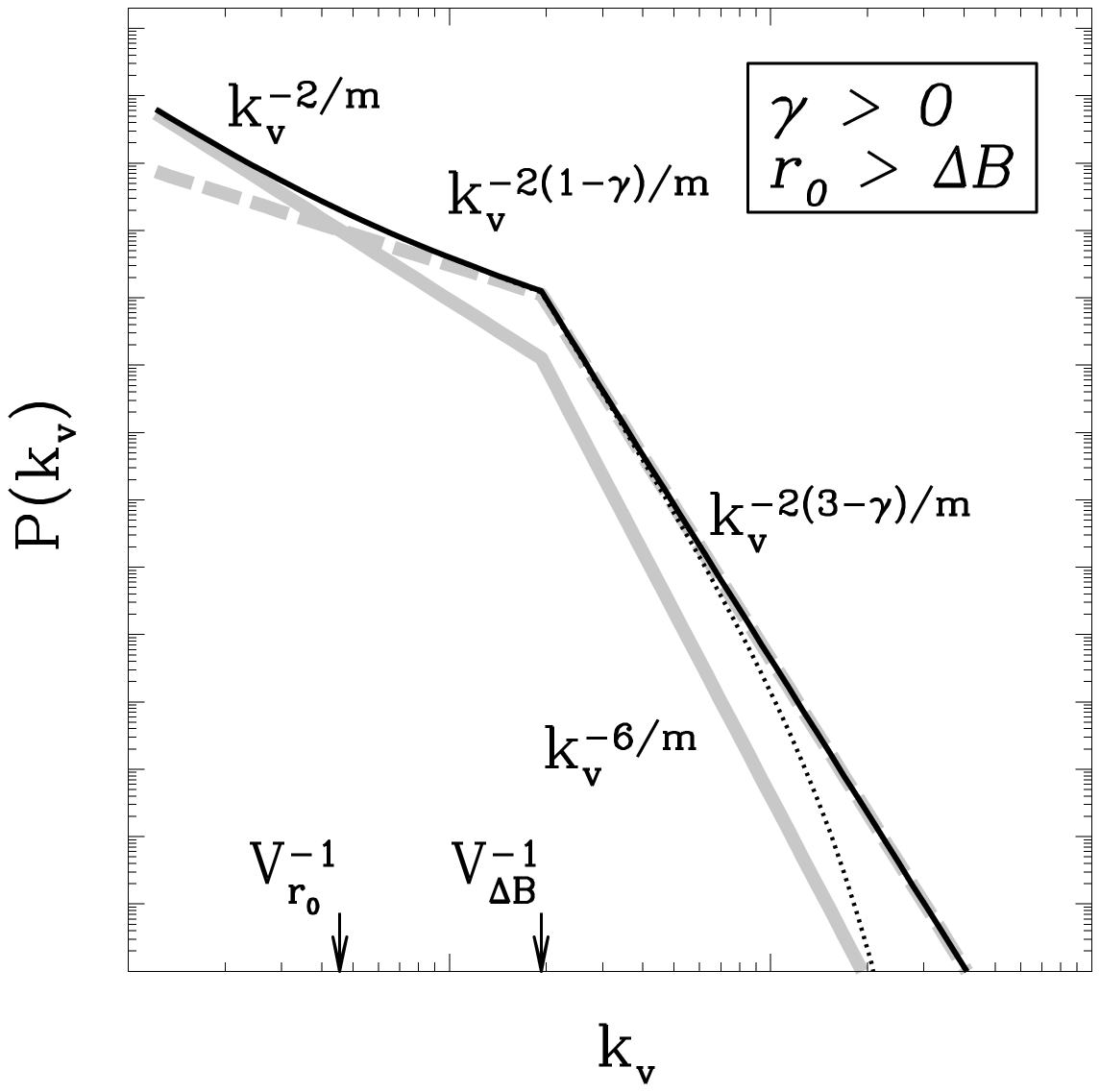}
\hfill
 \includegraphics[height=.3\textheight]{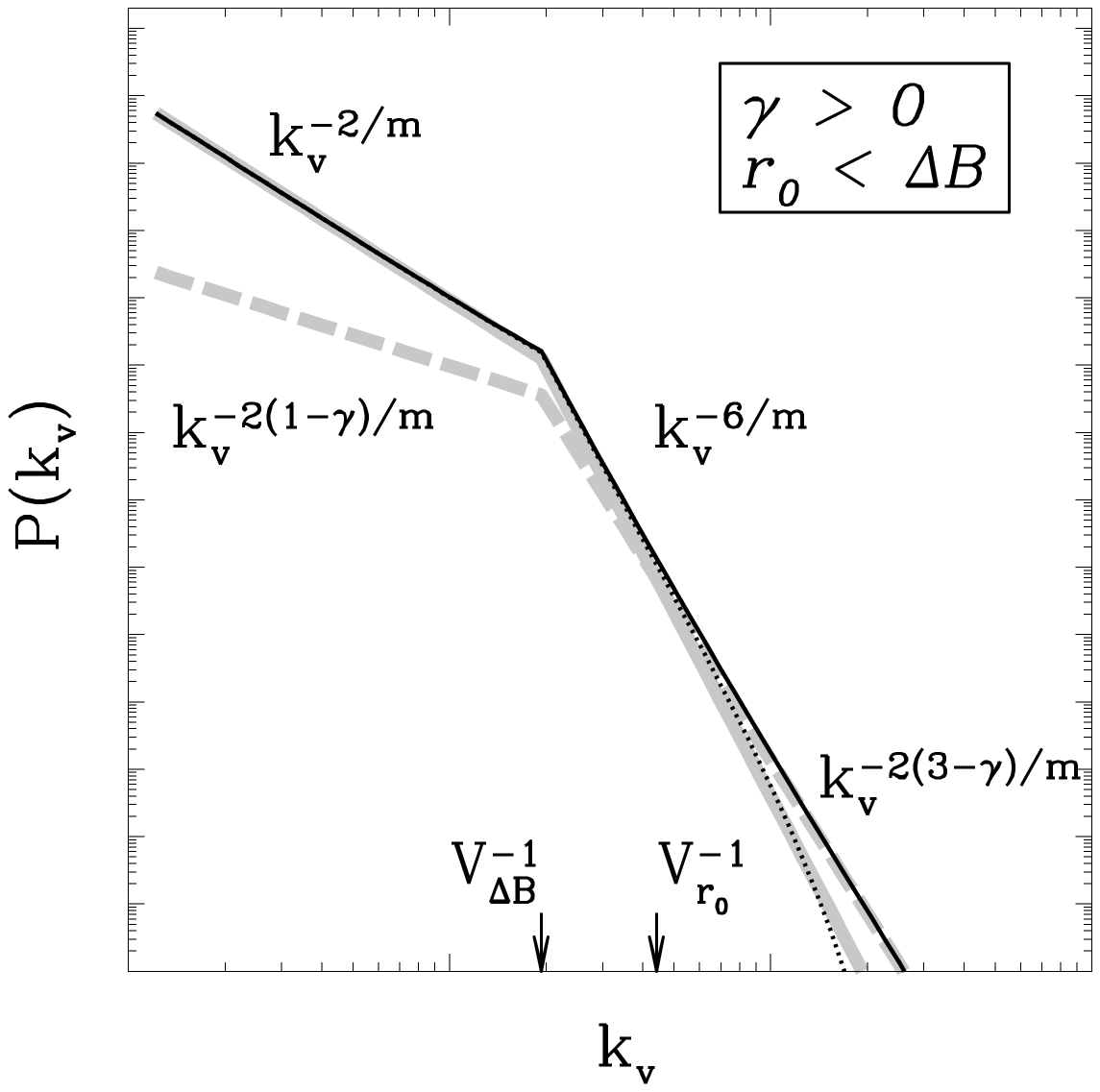}
\hfill
 \includegraphics[height=.3\textheight]{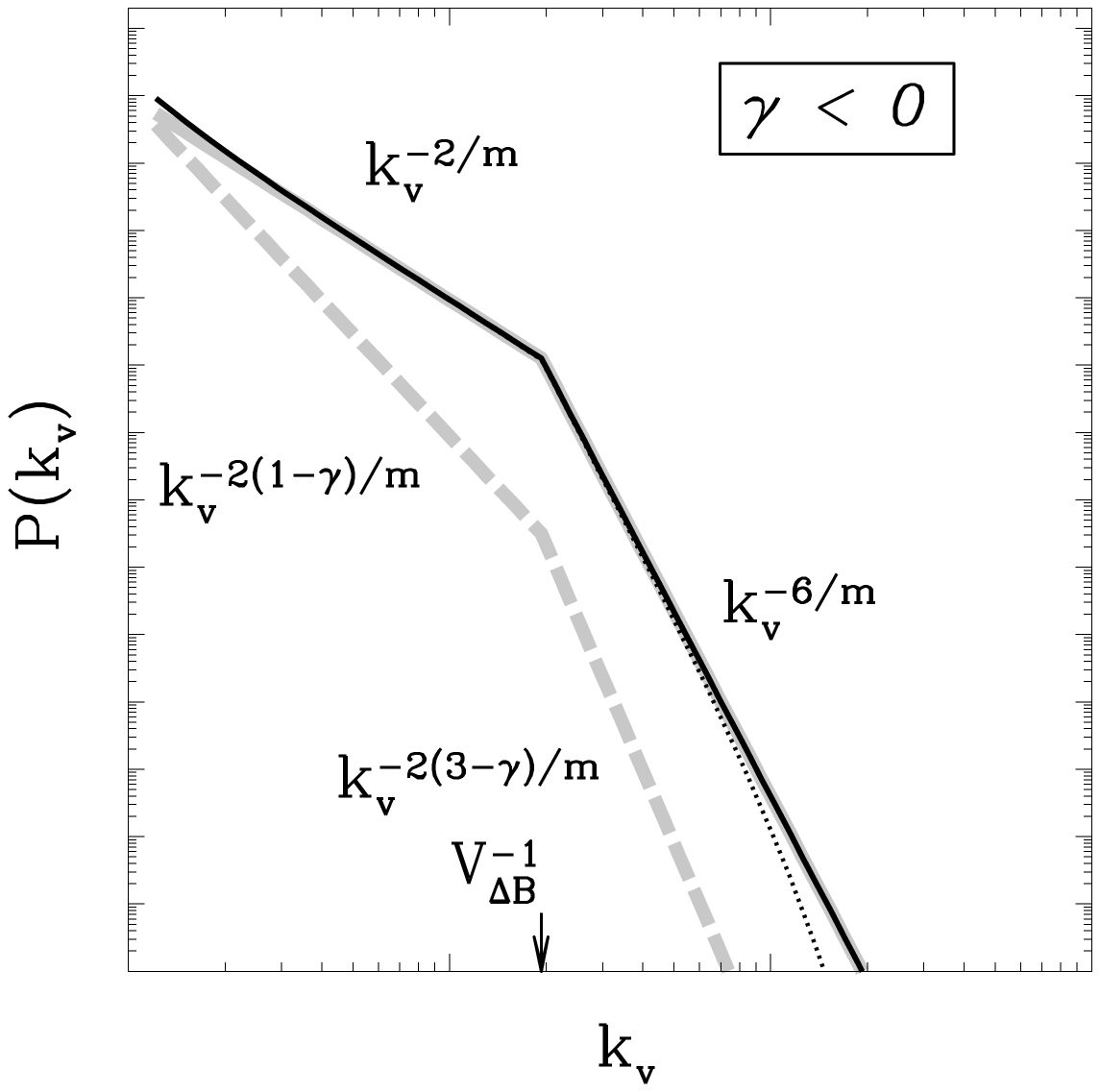}
}
  \caption{Particular cases for the VCS (from LP06).
In every panel light lines show contributions from the $\rho$-term
(density modified by velocity, dashed line) and $v$-term (pure velocity
effect, solid line) separately, while the dark solid line shows the combined
total VCS power spectrum. Thermal suppression of fluctuations
is shown by the dotted line. 
Labels below the dark solid lines mark the scaling of the subdominant
contributions.
For the {\it left} and {\it middle panels} the density
power spectrum is taken to be shallow, $\gamma > 0$.
The left panel corresponds to high amplitude of the density correlations,
$r_0 > \Delta B$, where density effects become dominant at relatively
long wavelengths for which the beam is narrow. In the middle panel,
the amplitude of density correlations is low $r_0 < \Delta B$ and they
dominate only the smallest scales which results in the intermediate steepening
of the VCS scaling. The {\it right panel} corresponds to the  steep density spectrum.
In this case the density contribution is always subdominant.
In this example the thermal scale is five times shorter than
the resolution scale $V_{\Delta B}$.}
\end{figure}

A realistic beam has a finite width, $\Delta B$. We remind the
reader that we deal with the case in which  
the turbulent volume extend along the line of sight $S$ is much
smaller than the distance to the volume. As the result,
the angular extend of the beam
is straightforwardly related to the physical scales that we deal with.
The VCS near a scale $k_v$ depends on whether the instrument  
resolves the correspondent spatial scale
$\left[k_v^2 D_z(S)\right]^{-1/m} S$. 
If this scale is resolved then $P_v(k_v) \propto k_v^{-2/m}$
and $P_{\rho}(k_v) \propto k_v^{-2(1-\gamma)/m}$. 
If the scale is not resolved then
$P_v(k_v) \propto k_v^{-6/m}$ and $P_\rho(k_v) \propto k_v^{-2(3-\gamma)/m}$. 
These results are presented in a compact form in Table~\ref{table:results}.
\begin{table}[h]
\begin{tabular}{lll} \hline\hline \\[-2mm]
Spectral~term & $\Delta B < S \left[k_v^2 D_z(S)\right]^{-\frac{1}{m}}$ &
$ \Delta B > S \left[k_v^2 D_z(S)\right]^{-\frac{1}{m}}$ \\[2mm]
\hline \\[-2mm]
$ P_\rho(k_v) $ & $ \propto\left(k_v D_z^{1/2}(S)\right)^{-2(1-\gamma)/m} $& 
$ \propto\left(k_v D_z^{1/2}(S)\right)^{-2(3-\gamma)/m} $ \\[2mm]
\hline \\[-2mm]
$ P_v(k_v) $ & $ \propto\left(k_v D_z^{1/2}(S)\right)^{-2/m} $  & 
$ \propto\left(k_v D_z^{1/2}(S)\right)^{-6/m} $ \\[2mm]
\end{tabular}
\caption{Scalings of VCS for shallow and steep densities (from LP06). 
To the linear scale $\Delta B$ corresponds 
the velocity
scale 
$
V_{\Delta B} \equiv \sqrt{D(S) (\Delta B/S)^m},$
equal to the magnitude of turbulent velocities at
the separation of a size $\Delta B$. It is not difficult to find that when
$
k_v^{-1} > V_{\Delta B} 
$
the beam is narrow, while on smaller scales its width is important.}
\label{table:results}
\end{table}
The transition from the low to the high resolution regimes happens as
the velocity scale under study gets comparable to the turbulent velocity
at the minimal spatially resolved scale. As the change of slope is the
velocity-induced effect, it is not surprising that the difference in
spectral indexes in the low and high resolution limit is $4/m$ for both
$P_v$ and $P_\rho$ terms, i.e it does not depend on the density\footnote{In
the situation where the available telescope resolution is not sufficient,
i.e. in the case of extragalactic turbulence research, the high spatial
resolution VCS can be obtained via studies of the absorption lines from
point sources.}.
This allows for separation of the velocity and density contributions.
For instance, Figure~2
illustrates that in the case of shallow density both the density and velocity 
spectra can be obtained. 
Potentially, procedures of extracting the 
information on the 3D turbulent
 density can be developed for the steep density case
as well. However, this requires careful accounting for errors that arise 
while the major contribution arising from velocity is subtracted from the 
data\footnote{Needless to say, when the turbulent object is resolved, the 
easiest way to obtain the density
spectral index is to study the integrated intensity maps, provided that the
absorption is negligible (see criteria for this in LP04).}.

The VCS is applicable to both emission and absorption studies of turbulence. A
numerical study in Chepurnov \& Lazarian (2006b) shows that having just several
lines of sight along which the absorption spectra is measured is sufficient for
recovering the underlying turbulent velocity spectrum (see Figure~3).
In the case of strong regular shear just one line of sight may be sufficient.

\begin{figure}
\hbox{
  \includegraphics[height=.3\textheight]{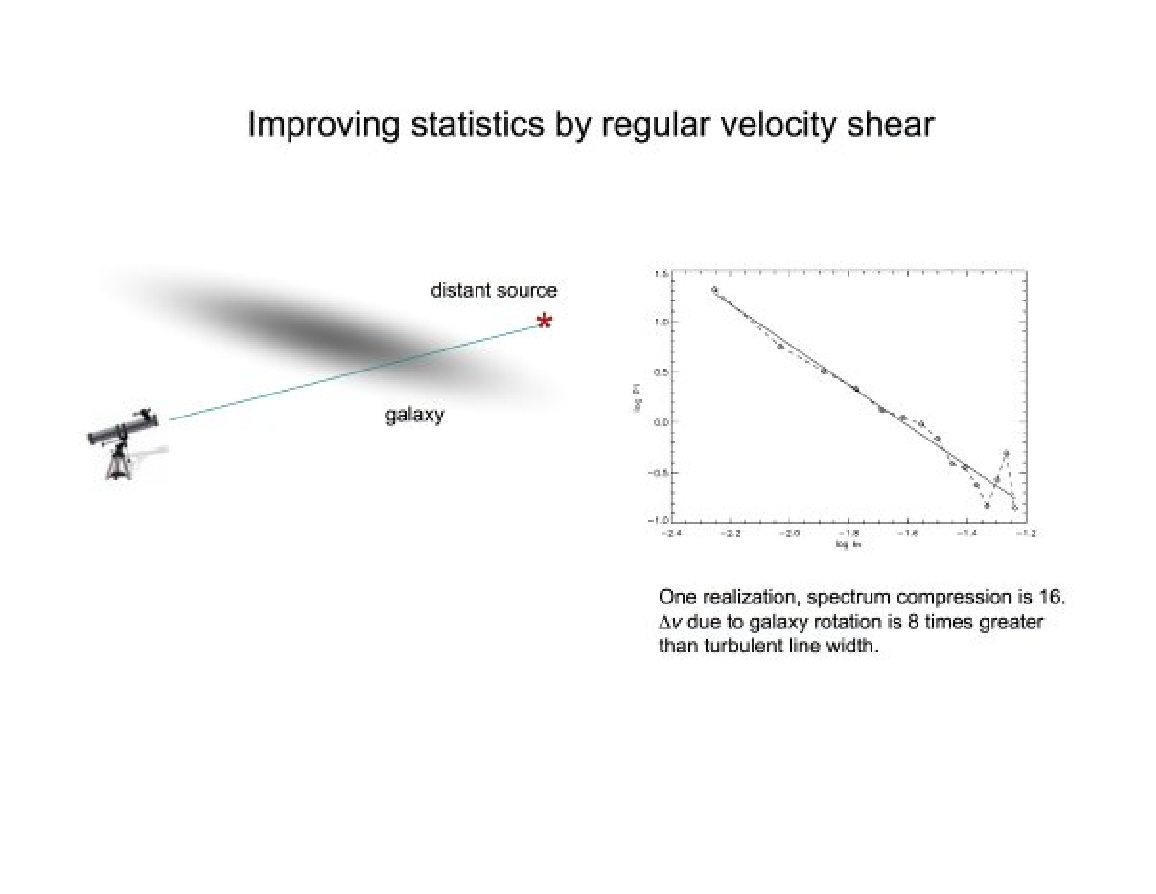}
\hfill
 \includegraphics[height=.3\textheight]{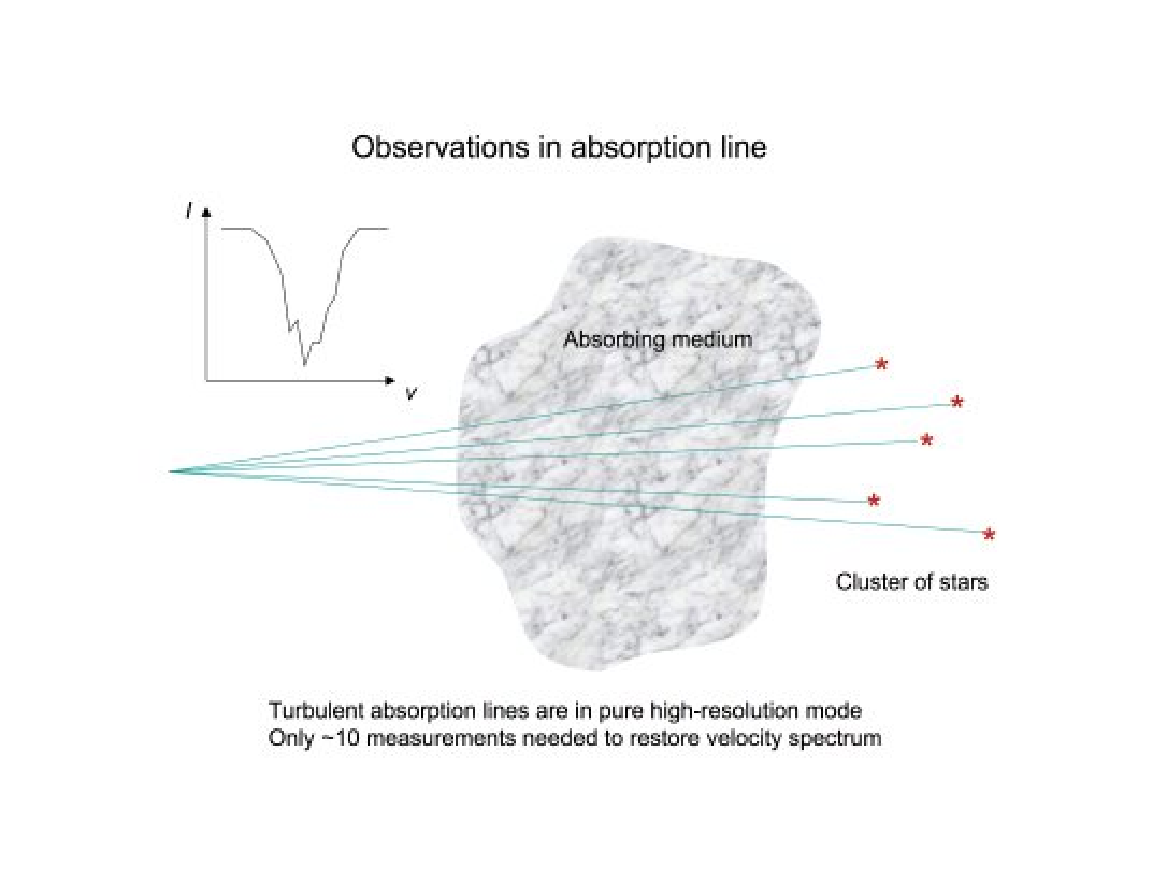}
}
  \caption{Illustration of VCS absorption studies of turbulence. {\it Left Panel}:  In 
case an 
external galaxy in which regular shear within the galaxy helps to improve the
statistics. {\it Right Panel}: For a cloud we showed that 
we can restore the turbulence
spectrum with spectroscopic measurements for less than 10 directions. 
From Chepurnov \& Lazarian 2006b}
\end{figure}

\section{Effects of Absorption}

The issues of absorption were worrisome for the researchers from the
very start of the research in the field (see Munch 1958). The erroneous
statements about the effects of absorption on the observed turbulence
statistics are widely spread in the literature (see discussion in LP04).

Using transitions that are less affected by absorption, e.g. HI,
may allow to avoid the problem. However, it looks regretful not
to use the wealth of spectroscopic data only because absorption
may be present. A study of absorption effects
in LP04 and LP06. For the VCA it was found that for sufficiently thin\footnote{The thermal 
broadening limits to what extend the slice can be thin. This means that in some cases
that the actual turbulent velocity spectrum may not be recoverable.} slices
the scalings obtained in the absence of absorption still hold
provided that the absorption on the scales under study is negligible.
The criterion for the absorption to be important is 
$\alpha^2 \langle (\rho_s({\bf X}, v_1)-\rho_s({\bf X},v_2))^2\rangle\sim 1$, which for $\gamma<0$
 results in the critical size of the slice thickness $V_c$ given by (LP06)
\begin{eqnarray}
V_{c}/D_z(S)^{1/2}
 &\approx& \left(\alpha \bar \rho_s\right)  ^{\frac{2m}{m-2}}, 
~~~~~~ m > 2/3 \label{eq:abs_width1}\nonumber \\
V_{c}/D_z(S)^{1/2}
 &\approx& \left(\alpha \bar \rho_s\right)^{-1}~,
~~~~~~ m < 2/3,
\label{eq:abs_width2}
\end{eqnarray}
where $\bar \rho_s$ is the mean PPV density.
The absorption is dominant for the slices thicker than $V_c$. The difference with the case of
$\gamma>0$ is that in the latter case one should also consider a possibility that density 
contribution can be important (see Table~2).
The criterion above coincides with one for the VCS, if we identify the critical
$k_v$ with $1/V_c$. If the resolution of the of the telescope is low, another limitation applies.
The resolved scale should be less than the critical spatial scale that arises from
the condition $\alpha^2 \langle (\rho_s({\bf X_1}, v)-\rho_s({\bf X_2},v))^2\rangle\sim 1$ which
for $\gamma<1$ results in $R_{c}/S \approx 
\left(\alpha \bar \rho_s\right)  ^{\frac{2}{m-2}}$ (LV06). If only  
scales larger than $R_c$ are resolved, the
information on turbulence is lost.

If integrated intensity of spectral lines is studied in the presence of
absorption non-trivial effects emerge. Indeed, for optically thin
medium the spectral line integration results in PPV intensity fluctuations that reflect
the density statistics. LP04 showed that this may not
be any more true for lines affected by absorption.
When velocity is dominant a very interesting regime for which
intensity fluctuations show universal behavior, i.e. the
power spectrum $P(K)\sim K^{-3}$  emerges.
When density is dominant (see Table~2),
the spectral index of intensity fluctuations in those two situations is the same
as in the case an optically thin cloud integrated through its volume. This  means
that for $\gamma>0$ in the range of parameter space defined by Table~2 
the measurements of intensity fluctuations of the integrated spectral
lines reflect the {\it actual} underlying density
spectrum in spite of the absorption effects.

\section{Comparison of Techniques}

Traditionally the techniques to study velocity
turbulence, e.g. velocity centroids or VCA, require the observations to spatially resolve the scale
of the turbulence under study\footnote{As it was discussed in LP00, the VCA
can be applied directly to the raw interferometric data, rather than to
images that require good coverage of all spatial frequencies. However, even
with interferometers, the application of the VCA to extragalactic objects is
restricted.}. This constrains the variety of astrophysical
objects where the turbulence can be studied.
In this way, the VCS,
is a unique tool that allows studies of astrophysical turbulence even when
the instrument does not resolve the turbulent fluctuations spatially. 

Our study of the effect of finite temperatures for the technique reveals that,
unlike the VCA, the temperature broadening does not prevent the turbulence
spectrum from being recovered from observations. Indeed, in VCA, gas
temperature acts in the same way as the width of a channel. Within the VCS
the term with temperature gets factorized and it influences the amplitude
of fluctuations (LP06). One can correct for this term\footnote{To do this, one may
attempt to fit for the temperature that would remove the exponential
fall off in the spectrum of fluctuations along the velocity coordinate
(Chepurnov \& Lazarian 2006a).}, which also allows for a new
way of estimating the interstellar gas temperature.

Another advantage of the VCS compared to the VCA is that it reveals the
spectrum of turbulence directly, while within the VCA the slope of the spectrum
should be inferred from varying the thickness of the channel. As the thermal
line width acts in a similar way as the channel thickness, additional care
(see LP04) should be exercised not to confuse the channel that is still
thick due to thermal velocity broadening with the channel that shows the 
thin slice asymptotics. A simultaneous
use of the VCA and the VCS makes the turbulence
spectrum identification more reliable.

The introduction of absorption in VCS and VCA brings about different results.
Within the analysis of velocity slices spectra (VCA) the absorption results
in new scalings for slices for which absorption is important. 
The turbulence spectral indexes 
can be recovered for the VCA within 
sufficiently thin slices, provided that the thickness of the slices
exceeds the thermal line width. For the VCS when absorption 
becomes important the spectra get exponentially damped. This
simplifies the interpretation of the
data.

Both VCA and VCS are applicable to studies of not only 
emission, but also absorption lines.  However, the necessity of
using extended emission sources limits the extent of possible VCA studies of
turbulence. This is not an issue for the VCS, for which absorption 
lines from {\it point sources} can be
used (see Figure~3). Interestingly enough, in this case the VCS asymptotics for the high
resolution limit should be used irrespectively of the actual beam size of the
instrument.   
Note, that the VCA and the VCS are applicable to lines for which emission
depends not only to first but also to the second power of density (see discussions
in LP04, LP06).

A more traditional approach to turbulence studies includes 
 velocity centroids, i.e.
$S(\mathbf{X})=\int v_{z}\ \rho_{s}(\mathbf{X},v_{z})\ {\rm d}v_{z}$,
where $\rho_{s}$ is the density of emitters in the  PPV space.\footnote{ Traditionally
the definition of centroids includes a normalization by the integral of $\rho_s$. This,
however does not substantially improve the statistics, but makes the
analytical treatment very involved (Lazarian \& Esquivel 2003).}.
Analytical expressions for structure functions\footnote{Expressions for the correlation 
functions are straightforwardly related to those of structure functions. 
The statistics of centroids using correlation functions was used in a later paper by 
Levier (2004).}   of centroids, i.e.
$ 
\left\langle \left[S(\mathbf{X_{1}})-S(\mathbf{X_{2}})\right]^{2}\right\rangle $
 were derived in Lazarian \& Esquivel (2003). In that paper a necessary criterion
for centroids to reflect the statistics of velocity was established.
 Esquivel \& Lazarian (2005) confirmed the utility of the criterion
and revealed that for MHD turbulence
simulations it holds for subsonic or slightly supersonic
turbulence (see also Ossenkopf et al. 2006).
This is in contrast to the VCA and the VCS  that provide reliable ways to study
supersonic turbulence.

VCA is related to the Spectral Correlation Functions 
(SCF) (see Padoan, Goodman \& Juvela 2003 and references therein).
The latter technique also deals statistics of velocity slices.
However if, ignoring the normalization, we write the informative part
of the SCF in our notations as 
$\langle[\int dv(\rho_s({\bf X_1}, v)-\rho_s({\bf X_2, v}))^2]^{1/2}\rangle$ we see
that we cannot commute the $()^{1/2}$ and $\langle \rangle$ operations.
This is very unfortunate, as if it had been possible, we the SCF would have corresponded
to $\xi_s^{1/2}(R, 0)$ (see Eq.~(\ref{ksicloud})) for which we have the analytical
theory. With all the wealth of data analyzed using the SCF (see Padoan et al. 2003) it is 
very tempting to investigate what sort of errors this mathematically incorrect commuting
 entails. Alternatively, it is appealing to reanalyze the data in Padoan et al. 2003
using the VCA.  It also seems interesting, although currently not feasible,
to find the connections between the VCA and VCS and another statistical
 tool, namely,
Principal Component Analysis  
(see Heyer \& Brunt 2004 and references therein).

In LP00, LP04, LP06 we used HI as an example of species to which the technique
is to be applied. 
Using heavier species that have lower thermal Doppler width of spectral lines
allows one to study turbulence up to smaller scales.  In addition, we would
like to stress that the VCS technique can be used at different wavelength. For
instance, the X-ray spectrometers with high spatial resolution can be used
to study of turbulence in hot plasma. In particular, the potential of VCS is
high for studies of turbulence in clusters of galaxies (cf.
Sunyaev et al. 2003 and references therein).  A simulated
example of such a study with the 
future mission Constellation X is provided in Lazarian (2006). 

Studies of turbulence in objects which are poorly resolved spatially is a
natural avenue for the VCS applications. Interestingly enough, in this case
one can combine the absorption line studies, which would provide the 
VCS for the pencil beam, i.e. for the high resolution, with the emission 
studies that would provide the VCS in the poor resolution limit. Potentially,
both velocity and density spectra can be obtained this way.

The importance of this work goes beyond the actual recovery of the
particular power-law indexes. First of all, the techniques can be generalized
to solve the inverse problem to recover non-power law turbulence spectra.
This may be important for studying turbulence at scales at which
either injection or dissipation of energy happens. Such studies are important
for identifying astrophysical sources and sinks of turbulent energy.
Second, studies of the VCS transition from low resolution to high resolution
regimes (see Figure~2) allows one to separate thermal and non-thermal contributions
to the line-widths as it is discussed in LP06. This could both
test the thermal correction that can be applied to extend
the power-law into sub-thermal velocity range (see also Chepurnov
\& Lazarian 2006b) and enable studies of temperature distribution
of the gas in atomic clouds (cf. Heiles \& Troland, 2003).

\section{Testing and Applying to Observations}

VCA predictions were tested in Lazarian et al. (2001), Esquivel et al. (2003)
and in Chepurnov \& Lazarian (2006b)
using synthetic maps obtained with 
simulated power-law data as well
as with numerical compressible MHD simulations.
Simulated data cubes allowed both density and velocity statistics
to be measured directly. Then these data cubes were
used to produce synthetic spectra which were analyzed using the
VCA. As the result, the velocity
and density statistics were successfully recovered\footnote{A recent
study in Chepurnov \& Lazarian (2006b) explains why an attempted testing of
the VCA in Miville-Deschenes et al. (2003) had inadequate resolution and therefore
 brought erroneous results.}.

The VCA has been applied to several data sets already. One of the first applications was in
Stanimirovic \& Lazarian (2001), where the technique was applied to the Small Magellanic Cloud (SMC) data.
The analysis revealed spectra of 3D velocity fluctuations roughly consistent with
the Kolmogorov scaling (a bit more shallow). LP00 argued that the Kolmogorov scaling
was expected for the magnetized turbulence appealing to
the Goldreich-Shridhar (1995) model. Esquivel et al. (2003)
used simulations of MHD turbulent flows to show that in spite
of the presence of anisotropy caused by magnetic field the
expected scaling of fluctuations is indeed Kolmogorov. Studies by
Cho \& Lazarian (2003) revealed that the Kolmogorov-type
scaling is also expected in the compressible MHD flows. This also
supports the conclusion in LP00 that the data in Green (1993)
is consistent with MHD turbulence scaling.

Muller et al. (2004) applied VCA to the Magellanic Bridge and discovered two structures with
distinctly different properties. If the spectrum of turbulence in the southern part is
similar to that in the SMC (velocity is roughly consistent with the Kolmogorov value and
the density somewhat shallower the Kolmogorov value), the northern part of the Magellanic Bridge
reveals completely different properties. If we interpret the results for the northern
part using Table~3 
the velocity power spectrum we get is very steep 
(index $\sim 4.2$ compared with Kolmogorov $\sim 3.7$),
while the density is shallow (index $\sim 2.8$). Whether these values should be interpreted
in terms of turbulence (some sort of shocks)
 or some other effects dominate is an open question that requires further
studies.

\begin{figure}
\hbox{
 \includegraphics[height=.3\textheight]{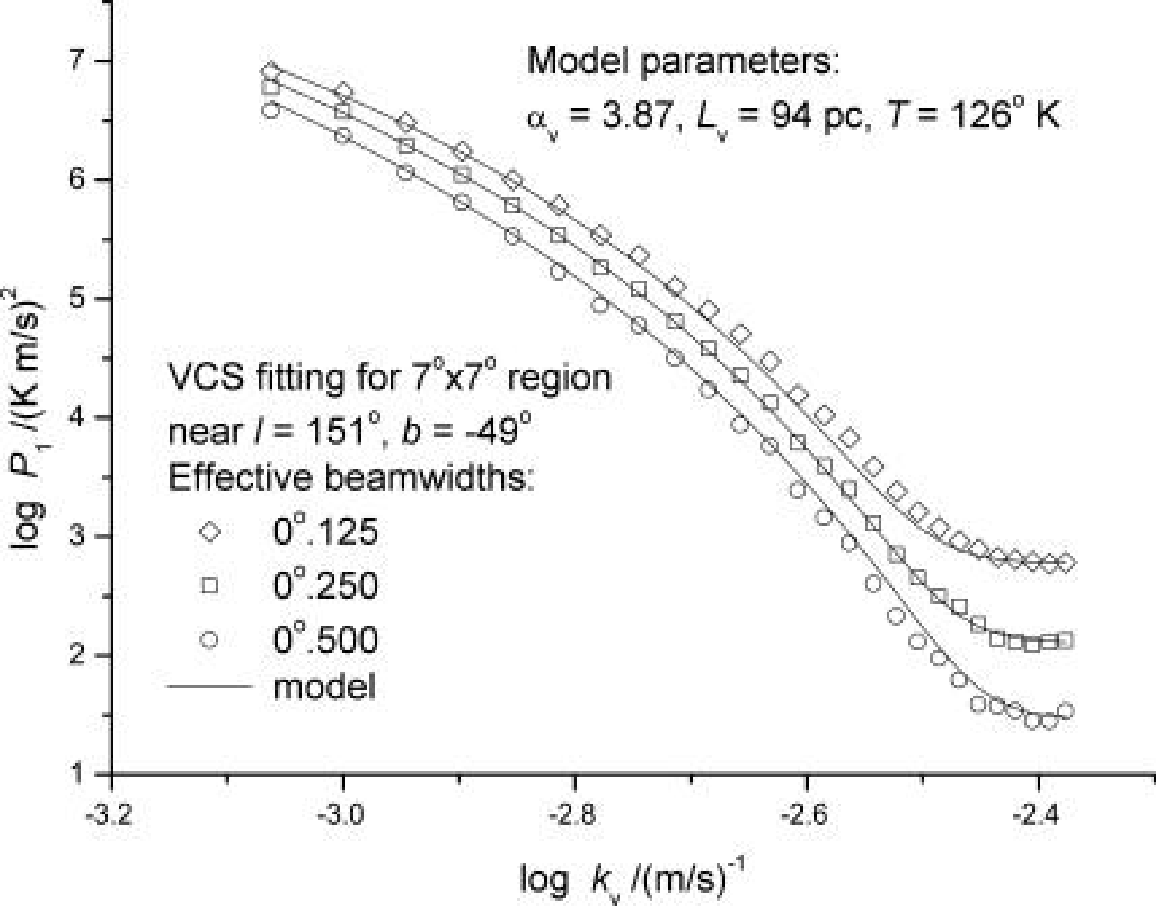}
}
  \caption{Application of the VCS to GALFA high latitude Arecibo data (by
Chepurnov \& Lazarian). The spatial resolution of the maps was decreased to illustrate
the VCS in both high and low resolution regimes. $\epsilon$ here is the power spectrum index, which for the Kolmogorov
turbulence is $11/3$. The measured $\epsilon$ is in the range
$[3.52; 3.57]$.}
\end{figure}

Studies of turbulence are more complicated for the inner parts of the Galaxy,
where (a) two distinct regions at different distances from the observer
contribute to the emissivity for a given velocity and (b) effects of
the absorption are important. However, the analysis in Dickey et al. (2001)
showed that some progress may be made even in those unfavorable
circumstances. Dickey et al. (2001) found the steepening
 of the spectral index with the increase of the velocity slice thickness.
They also observed the spectral index for strongly absorbing direction
approached $-3$ in accordance with the conclusions in LP04.
Note,  that  the effects of optical depths may explain
some other case when the spectral index stayed the same, e.g. -3, while the thickness of the slice
was varying (see Kralil et al. 2006). Incidently, this situation
 can be confused with the situation when 
the fluctuations arise from density only (see Begum et al. 2006).

21-cm absorption provides another way of probing turbulence on small
scales. The absorption depends on the density to temperature ratio
$\rho/T$, rather than to $\rho$ as in the case of emission\footnote{
In the case of an isobaric medium the product of
density and temperature are constant and the problem is similar to studies of 
transitions for which the emissivity is proportional to $\rho^2$ that we discussed
earlier.}. However,
 in terms of the VCA this change is not important and we still expect to
see emissivity index steepening as velocity slice thickness increases,
provided that velocity effects are present. In view of
this, results of Deshpande et al. (2001), who did not see such steepening,
can be interpreted as the evidence of the viscous suppression of
turbulence on the scales less than 1~pc. The fluctuations in this
case should be due to density and their shallow spectrum $\sim k^{-2.8}$ may
be related to the damped magnetic structures below the viscous
cutoff (see Lazarian, Vishniac, \& Cho 2004).
This may be  also a consequence of the shallow 
density spectrum in compressible MHD 
(see Beresnyak, Lazarian \& Cho 2005).

Historically, the CO data was analyzed after integration over the entire emission line. 
Stutzki et al. (1998) presented the power spectra of  $^{12}$CO and $^{13}$CO 
fluctuations obtained via integrating the intensity over the entire emission line
for L1512 molecular cloud. 
Counter-intuitively, Stutzki et al. (1998) found for both isotopes the power spectrum
with a similar spectral index. According to LP04 this may correspond to optically thick
assymptotics (i.e. the integration range of velocities is larger than $V_c$ (see Eq.~{\ref{eq:abs_width2})).
 If the velocity fluctuations dominate, the expected index is universal
and equal to $-3$ (meaning $K^{-3}$),
if the density fluctuations dominate (see Table~2) the expected index is $-3+\gamma$
(meaning $K^{-3+\gamma}$).
The index measured in Stutzki et al. (1998) is $\sim 2.8$, which may either correspond to
$-3$ within the experimental errors, or more likely indicate the that $\gamma\approx 0.2$,
i.e. the density spectrum is shallow. The latter possibility is indirectly supported by
$^{18}CO$ measurement for L1551 cloud in Swift (2006), used the VCA (observing
the changes of the channel map spectral index while changing the
velocity slice thickness) and obtained the shallow density spectrum with $\gamma \approx 0.2$,
while his measured velocity spectrum was approximately Kolmogorov (the index is $-3.72$).
Padoan et al. (2006) both successfully tested the VCA with high resolution numerical simulations
that included radiative transfer and applied the technique to Five College Radioastronomy Observatory
(FCRAO) survey of the Perseus molecular cloud complex. He obtained the velocity index around $-3.81$.

VCS predictions were successfully tested in Chepurnov \& Lazarian (2006b). 
In Chepurnov \& Lazarian (in preparation) the technique was applied to
the Arecibo high latitude galactic data. Figure~4 illustrates the results
obtained for maps with high spectral resolution as well as for the spatially smoothed
maps with lower spatial resolution. This allowed to study both the high and
low resolution regimes of the VCS. The results obtained for the aforementioned regimes
are similar, namely, the power spectrum indexes are around $-3.87$, which is somewhat steeper than the
Kolmogorov index. 



The practical application of the VCA and the VCS has only started. Nevertheless, it has 
already provided some intriguing results
and proved to be a promising way of using the wealth of spectroscopic surveys for
studies of astrophysical turbulence. 


{\bf Acknowledgments} I thank Alexey Chepurnov, Dmitry Pogosyan and Snezana Stanimirovic
for their input.  AL research is supported by
by NSF grant AST 0307869 and the NSF Center for Magnetic Self Organization in
Laboratory and Astrophysical Plasmas.


\end{document}